\documentclass[aps,pra,reprint, superscriptaddress]{revtex4-1}

\usepackage{epsfig,graphics,graphicx}
\usepackage[export]{adjustbox}
\usepackage{dcolumn}
\usepackage{bm}
\usepackage{color}
\usepackage{amsmath,amssymb,amsthm}
\usepackage{centernot}
\usepackage{bbold}
\usepackage{mathtools}
\usepackage{fullpage}
\usepackage{units}
\usepackage[usenames,dvipsnames]{xcolor}
\usepackage{braket}
\usepackage{indentfirst}

\usepackage[utf8]{inputenc}
\usepackage[T1]{fontenc}

\usepackage{mathpazo}
\usepackage{dsfont}

\usepackage{mathptmx}
\usepackage{amssymb}
\usepackage{amsmath}
\usepackage{latexsym}
\usepackage{amsfonts}

\usepackage{hyperref}
\hypersetup{pdfpagemode=UseNone}

\newtheorem{theorem}{Theorem}
\newtheorem{definition}[theorem]{Definition}

\newtheorem{lemma}[theorem]{Lemma}
\newtheorem{proposition}[theorem]{Proposition}

\newcounter{rem}
\newcommand{\bracket}[2]{\ensuremath{\left\langle#1 \vphantom{#2}\right| \left. #2 \vphantom{#1}\right\rangle}}

\newcommand{\paren}[1]{\ensuremath{\left(#1\right)}}

\def\>{\rangle}
\def\<{\langle}

\newcommand{\Mod}[1]{\ensuremath{\left|#1\right|}}
\renewcommand{\rho}{\varrho}

\def\textbf#1{{\bf #1}}

\def\beq{\begin{equation}}
\def\eeq{\end{equation}}
\def\beqa{\begin{eqnarray}}
\def\eeqa{\end{eqnarray}}
\def\eea{\end{array}}
\def\bea{\begin{array}}
\newcommand{\bei}{\begin{itemize}}
	\newcommand{\eei}{\end{itemize}}
\newcommand{\bee}{\begin{enumerate}}
	\newcommand{\eee}{\end{enumerate}}
\def\bep{\begin{proposition}}
	\def\eep{\end{proposition}}
\def\bel{\begin{lemma}}
	\def\eel{\end{lemma}}
\def\bet{\begin{theorem}}
	\def\eet{\end{theorem}}
\def\bed{\begin{definition}}
	\def\eed{\end{definition}}

\usepackage{soul}

\begin{document}

	\title{Multiparticle quantum walk with a gas-like interaction}
	
	\author{Pedro C. S. Costa}
	\affiliation{Mackenzie Presbyterian University, \\  Rua da Consola\c{c}\~{a}o 896, 
		São Paulo, SP - Brazil}
	\author{Fernando de Melo}
	\affiliation{Brazilian Center for Research in Physics (CBPF), \\ Rua Dr. Xavier Sigaud, 150, Rio de Janeiro, RJ, Brazil}
	\author{Renato Portugal}
	\affiliation{National Laboratory of Scientific Computing (LNCC), \\ Av. Getúlio Vargas, 333, Petrópolis, RJ, 25680-320, Brazil}

	\date{\today}
	
	
	\begin{abstract}
		We analyze the dynamics of multiparticle discrete-time quantum walk on the two-dimensional lattice, with an interaction inspired on a classical model for gas collision, called HPP model. In this classical model, the direction of motion changes only when the particles collide head-on, preserving momentum and energy. In our quantum model, the dynamics is driven by the usual quantum-walk evolution operator if the particles are on different nodes, and is driven by the HPP rules if the particles are in the same node, linearly extended for superpositions. Using this new form of evolution operator, we numerically analyze three physical quantities for the two-walker case: The probability distribution of the position of one walker, the standard deviation of the position of one walker, and the entanglement between the walkers as a function of the number of steps. The numerical analysis implies that the entanglement between the walkers as a function of the number of steps initially increases and quickly tends to a constant value, which depends on the initial condition.
		We compare the results obtained using the HPP interaction with the equivalent ones using the phase interaction, which is based on an evolution operator that inverts the sign of the coin operator if the walkers are in the same position.
	\end{abstract}

	\pacs{03.65.Ud, 03.65.Yz, 03.67.Bg, 42.50.Pq}
	
	\maketitle
	

	
	\section{Introduction}\label{sec:intro}
	
	In early 1970's, Hardy, Pomeau, and de Pazzis introduced a cellular automata model to describe a classical lattice gas of colliding particles~\cite{HPP73,HPP76}. Their model, known as the HPP model, is arguably the simplest description of a gas, with the allowed positions for the particles being the nodes of a two-dimensional square lattice. The dynamics, i.e.~the updating rule, is split into two parts called collision and streaming (or propagation). Particles on different sites propagate along the lattice with constant velocity. If two particles collide at a given site, their direction of motion is changed.  The collisions are described by a local rule that conserves the total momentum and the total number of particles. This first type of lattice-gas cellular automata, also called partitioned cellular automata, was highly improved and nowadays its variations are  used to simulate complex fluid dynamics, such as those obeying the Navier-Stokes differential equation, and are useful for modeling diffusion, thermalization processes, and  microscopic reversibility~\cite{Wol00,TM87}. Quantum versions of the HPP model have already been analyzed, for instance, in Refs.~\cite{SL13,love}.

	Quantum walks~\cite{ADZ93} are the quantum counterpart of random walks and have been successfully used to build quantum algorithms for problems such as element distinctness~\cite{Amb07,Por18}, and spatial search~\cite{SKW03,Por18Book}. Quantum walks are universal for quantum computation~\cite{CGW13,LCETK10}, and as such they are used to simulate quantum dynamics  as those governed by the Dirac equation~\cite{DMPT16}, and also neutrino oscillations~\cite{MP16}. Two-particle quantum walks were applied to the graph isomorphism problem~\cite{GFZJC10,BW11}, and also to show that an interacting quantum walk dynamics might lead to the formation of a stable compound state~\cite{AAMSWW12}. In these references, as well as in others~\cite{BW11,OPSB06,SolTwoW,RLW15,WL16}, the interaction is described by an operation that changes the relative phase between the walkers if they are on the same node.

	In this work, we introduce a multiparticle coined quantum walk with an interaction scheme based on the classical HPP model. We focus on the two-dimensional lattice, but the model can be straightforwardly generalized to $N$-dimensional lattices, with $N>2$. In quantum walk models, if $\ket{\psi(t)}$ describes the state of the system after $t$ steps, $U\ket{\psi(t)}$ describes the state of the system after $t+1$ steps, where $U$ is the evolution operator.  In our model, a basis state of the computational basis is represented by $\ket{c_1,x_1,y_1}\ket{c_2,x_2,y_2}$, where $\ket{c_1,x_1,y_1}$ \big($\ket{c_2,x_2,y_2}$\big) is the state of the first (second) particle, $c_1$ ($c_2$) is the coin value, and $(x,y)$ is a lattice node. The action of $U$ on a basis state is equal to the action of the evolution operator of two independent quantum walks if $(x_1,y_1)\neq (x_2,y_2)$ and it is equal to the action of  an operator designed based on the HPP rules, if $(x_1,y_1)= (x_2,y_2)$. The action of this operator takes into account the direction of motion of the walkers. 
	We impose an exclusion principle so that there are at most four particles in a node (in the two-dimensional case) with differing momentum direction. The particles interact when there is a head-on collision, which changes the direction of motion of each particle. As expected in quantum-walk models, there is no long range interaction, since the graph edges designate which pairs of nodes have a direct connection. The model works both for distinguishable and indistinguishable particles.

	Due to computational limitations, we address the case of two walkers on the two-dimensional lattice.
	To understand the quantum walk dynamics, we numerically analyze and describe three quantities: The probability distribution of the position of one walker, the standard deviation of the position of one walker, and the von Neumann entanglement entropy of the density matrix of one walker as a function of the number of steps. The standard deviation is a linear function with respect to the number of steps and the slopes are determined by the choice of the initial state. On the other hand, the entanglement as a function of the number of steps tends to a constant value, which also depends on the initial state. 
	These numerical results indicate that the effective strength of the interaction between the particles decreases as a function of time.
	We compare the results obtained using the HPP interaction with the equivalent ones using the phase interaction when the phase is $\pi$, that is, the sign of the coin operator inverts if the walkers are in the same position.

	
	The main motivation in this work is to contribute with the development of a quantum-walk-based model that might reproduce the behavior of quantum gases. As a second motivation, we are interested in applying multiparticle quantum walks for searching a marked node in the two-dimensional lattice. Our hope is based on the fact that we can improve the spreading rate of probability distribution of the position of a walker by choosing an appropriate initial condition. 
	
	The paper is organized as follows. In Section~\ref{sec:coinedM}, we review the classical HPP model and the coined quantum walk on the two-dimensional lattice. In Section~\ref{sec:model}, we introduce our model by describing the evolution operator of the multiparticle quantum walk with the HPP interaction. In Section~\ref{sec:numerics}, we present the numerical results with the focus on three quantities: the probability distribution, the standard deviation, and the entanglement. Finally, in Section~\ref{sec:conc}, we present our final considerations and discuss extensions of our model.

	\section{Preliminaries}
	\label{sec:coinedM}

	\subsection{Classical HPP model}
\begin{figure}
		\includegraphics[trim=30 30 30 30,clip,scale=0.35]{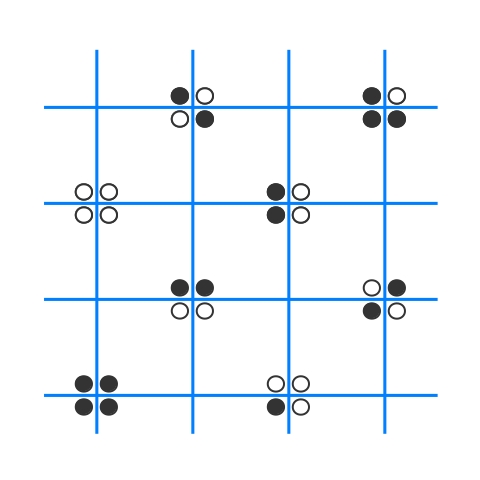}
		\caption{ \label{fig:HPPsites} Part of a two-dimensional lattice displaying 16 nodes and 4 cells per node. A node has label $(x,y)$ such that $x+y$ is even. A square necessarily contains 2 cells. A full cell represents one particle and a empty cell represents no particle.}
	\end{figure}

	In this subsection, we review the classical HPP model proposed by Hardy, Pomeau, and  de Pazzis~\cite{HPP73,HPP76,Wol00,CD98}. This model is one of the simplest examples of lattice-gas cellular automata. The setup is as follows. The positions of the particles are cells on a two-dimensional square lattice, whose nodes are described by $(x,y)$ such that $x+y$ is even. Each node has four cells (empty or full circles) as shown in Fig.~\ref{fig:HPPsites}. A full cell means the presence of one particle and an empty cell means the absence of particle.  
	
	The evolution operator is the composition of two operators, called collision $C$ and streaming $S$ (or propagation). The dynamics consists in applying the evolution operator over and over. The streaming operator moves the particle to its opposite cell, with the particle not leaving the square. If two particles are in the same square, they interchange positions. This description works both for the distinguishable and indistinguishable cases. Note that $S^2=I$.
	
\begin{figure}
		\includegraphics[trim=60 60 60 60,clip,scale=0.25]{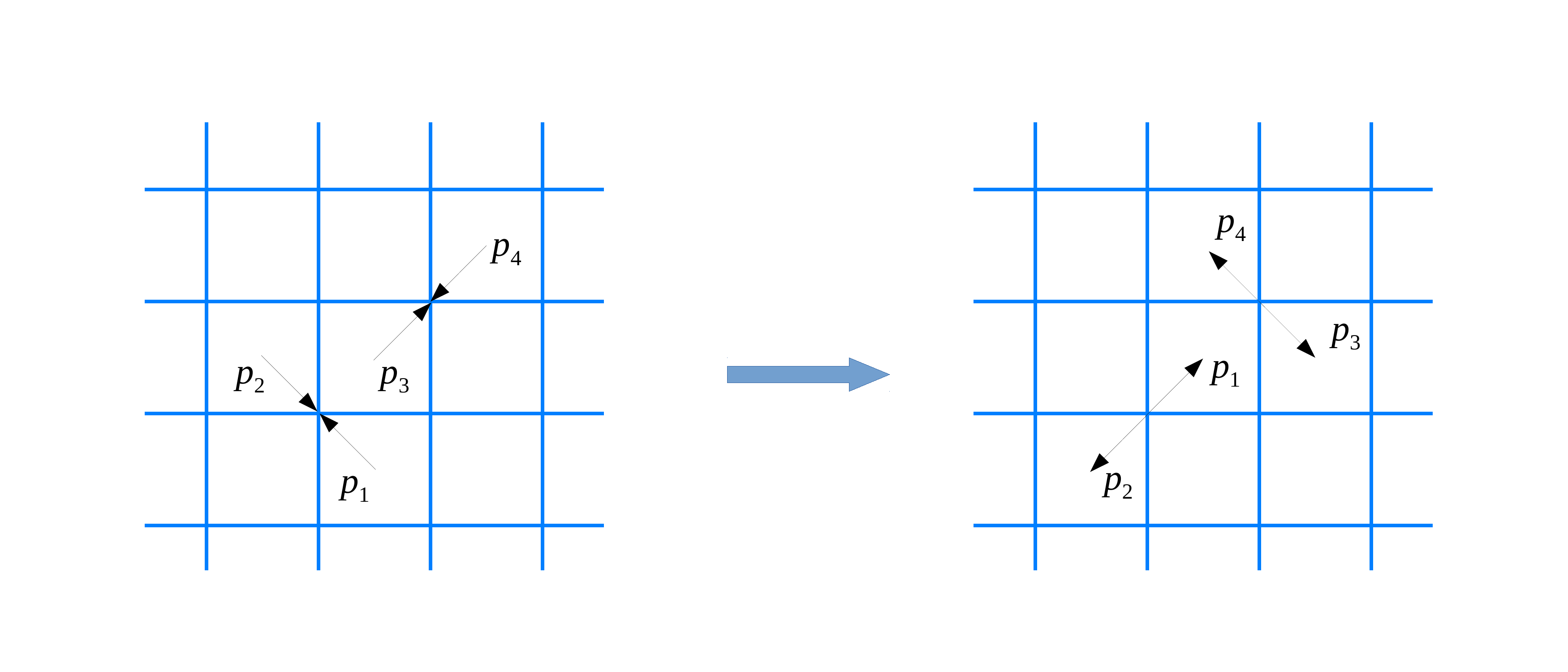}
		\caption{ \label{fig:collision} Two head-on collisions, where the particles are represented by the arrows (two cells are full and two cells are empty). The right-hand lattice shows the direction the particles take after the collision. In all other cases, the particles keep the same direction.}
	\end{figure}

	The collision operator $C$ acts on particles at the same node. There are at most 4 particles at a node, described by a block of 4 cells on 4 neighboring different squares (exclusion principle). The action of the collision operator changes the configuration of the block of 4 cells according to the following rules:
	\begin{align*}
	\mbox{}_{\circ}^{\circ}\mbox{}_{\circ}^{\circ} \,\,\,\,\longrightarrow\,\,\,\,  \mbox{}_{\circ}^{\circ}\mbox{}_{\circ}^{\circ} \hspace{0.7cm}
	\mbox{}_{\bullet}^{\circ}\mbox{}_{\circ}^{\circ} \,\,\,\,\longleftrightarrow\,\,\,\,  \mbox{}_{\circ}^{\circ}\mbox{}_{\circ}^{\bullet} \hspace{0.7cm}
	\mbox{}_{\circ}^{\bullet}\mbox{}_{\circ}^{\circ} \,\,\,\,\longleftrightarrow\,\,\,\,  \mbox{}_{\circ}^{\circ}\mbox{}_{\bullet}^{\circ} \\ \\
	\mbox{}_{\bullet}^{\circ}\mbox{}_{\bullet}^{\circ} \,\,\,\,\longleftrightarrow\,\,\,\,  \mbox{}_{\circ}^{\bullet}\mbox{}_{\circ}^{\bullet} \hspace{0.7cm}
	\mbox{}_{\bullet}^{\bullet}\mbox{}_{\circ}^{\circ} \,\,\,\,\longleftrightarrow\,\,\,\,  \mbox{}_{\circ}^{\circ}\mbox{}_{\bullet}^{\bullet}  \hspace{0.7cm}
	\mbox{}_{\bullet}^{\circ}\mbox{}_{\circ}^{\bullet} \,\,\,\,\longleftrightarrow\,\,\,\,  \mbox{}_{\circ}^{\bullet}\mbox{}_{\bullet}^{\circ} \\ \\
	\mbox{}_{\bullet}^{\circ}\mbox{}_{\bullet}^{\bullet} \,\,\,\,\longleftrightarrow\,\,\,\,  \mbox{}_{\bullet}^{\bullet}\mbox{}_{\circ}^{\bullet}  \hspace{0.7cm}
	\mbox{}_{\circ}^{\bullet}\mbox{}_{\bullet}^{\bullet} \,\,\,\,\longleftrightarrow\,\,\,\,  \mbox{}_{\bullet}^{\bullet}\mbox{}_{\bullet}^{\circ}  \hspace{0.7cm}
	\mbox{}_{\bullet}^{\bullet}\mbox{}_{\bullet}^{\bullet} \,\,\,\,\longrightarrow\,\,\,\,  \mbox{}_{\bullet}^{\bullet}\mbox{}_{\bullet}^{\bullet} 
	\end{align*}
	The rules $\mbox{}_{\bullet}^{\circ}\mbox{}_{\circ}^{\bullet} \,\longleftrightarrow\, \mbox{}_{\circ}^{\bullet}\mbox{}_{\bullet}^{\circ}$ represent head-on collisions. After a head-on collision along the secondary diagonal, the particles take the main diagonal and vice versa, as described in Fig.~\ref{fig:collision}. For distinguishable particles, we have to track each particle. In this case, the full cells must be labeled, for instance, the particles of the rules $\mbox{}_{\bullet_1}^{\circ}\mbox{}_{\circ}^{\bullet_2} \,\longleftrightarrow \, \mbox{}_{\circ}^{\bullet_2}\mbox{}_{\bullet_1}^{\circ}$, which inverts the diagonal directions, have labels 1 and 2. The action of the collision operator necessarily changes the positions of the cells in a block in the distinguishable case, including the case of four full cells, which is given by $\mbox{}_{\bullet_2}^{\bullet_1}\mbox{}_{\bullet_4}^{\bullet_3} \,\,\,\,\longrightarrow\,\,\,\,  \mbox{}_{\bullet_4}^{\bullet_3}\mbox{}_{\bullet_2}^{\bullet_1}$.  If it is not a head-on collision, diagonal cells in the same block are interchanged. All rules preserve time-reversibility. Note that the collision operator has the following properties: 1)~$C^2=I$ and 2)~momentum and energy (number of particles) are conserved.

	\subsection{Quantum walks on two-dimensional lattices}
\begin{figure}
		\includegraphics[trim=30 30 85 85,clip,scale=0.4]{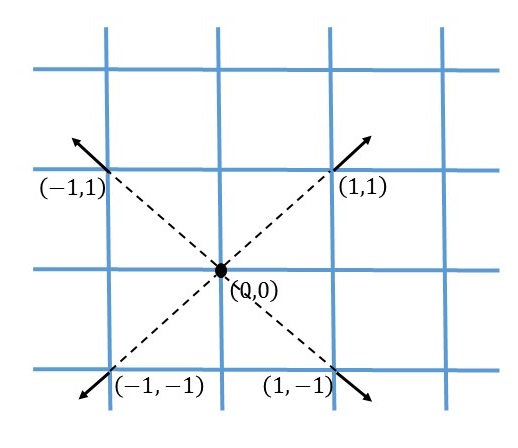}
		\caption{{ \label{fig:LQHPP} If the position of a walker is node $(0,0)$, there are four nodes it can go after the action of the shift operator. The nodes are $(1,1)$, $(-1,1)$, $(-1,-1)$, and (1,-1) associated with the coin states $(0,0)$, $(1,0)$, $(1,1)$, and $(0,1)$, respectively. The walker never steps on nodes $(x,y)$ such that $x+y$ is odd.}}
	\end{figure}

	In this subsection, we describe the dynamics of a single quantum walker on a two-dimensional square lattice. In the coined model, the Hilbert space is the tensor product of the coin and position spaces, as follows 
	\begin{equation}
	\label{eq:Hilbert_coined}
	\mathcal{H}_{C}\otimes\mathcal{H}_{P},
	\end{equation}
	where $\mathcal{H}_{P}$ is spanned by the set of states $\left\{ \ket{x,y} :x,y\in\mathbb{Z}\right\}$ when the lattice is infinite and by $\left\{ \ket{x,y} :x,y=0,1 \ldots, N-1\right\}$ when the lattice has cyclic boundary conditions, where $N$ is a positive integer and $N^2$ is the number of nodes. The coin space $\mathcal{H}_C$ has four dimensions and is spanned by $\left\{ \ket{c_{x},c_{y}} : c_{x},c_{y}\in \{0,1\} \right\}$. The action of the shift operator on a basis state is given by
	\begin{equation}
	\label{eq:shift}
	S\ket{c_{x},c_{y}}\ket{x,y} =\ket{c_{x},c_{y}}\ket{x+(-1)^{c_x},y+(-1)^{c_y}}.
	\end{equation}
	Each state $ \ket{c_{x},c_{y}}$ is associated with a direction of motion in the following way (see Fig.~\ref{fig:LQHPP}): 
	\begin{eqnarray}
	\label{eq:moviment_directions}
	\ket{00} \nearrow \label{00}, \,\,\,\,
	\ket{01} \searrow\label{01}, \,\,\,\,
	\ket{10} \nwarrow\label{10}, \,\,\,\,
	\ket{11} \swarrow\label{11}.
	\end{eqnarray}
	For instance, if $(c_x,c_y)=(0,0)$, the values of $x$ and $y$ are incremented by one unit under the action of $S$ and a walker at $(0,0)$ goes to $(1,1)$, that is, it goes northeast (see Fig.~\ref{fig:LQHPP}). A state of the coins describes the \textit{direction of the momentum} of the walkers.

	The generic state of the walker at time $t$ is 
	\begin{equation}
	\label{eq:gen2D}
	\ket{\psi\paren{t}} =\sum_{c_{x},c_{y}=0}^{1}\sum_{x,y}\psi_{c_{x}c_{y}}\paren{x,y,t}\ket{c_{x},c_{y}}\ket{x,y}, 
	\end{equation}
	where $\psi_{c_{x}c_{y}}\paren{x,y,t}$ are the amplitudes of the
	walker's state, which obey the normalization condition
	\[
	\sum_{c_{x},c_{y}}\sum_{x,y}\left|\psi_{c_{x},c_{y}}\paren{x,y,t}\right|^{2}=1,
	\] 
	for all integer time $t$.
	
	The evolution operator is
	\begin{equation}
	\label{eq:coin_evolution}
	U=S\cdot\left(C\otimes I\right),
	\end{equation}
	where $I$ is the identity operator acting on $\mathcal{H}_{P}$, $S$ is the shift operator, and $C$ is a four-dimensional coin. The most used coin is the Grover coin~\cite{Por18Book} given by 
	\begin{equation}
	\label{eq:grover_coin}
	C=\frac{1}{2}\begin{pmatrix}-1 & 1 & 1 & 1\\
	1 & -1 & 1 & 1\\
	1 & 1 & -1 & 1\\
	1 & 1 & 1 & -1
	\end{pmatrix}.
	\end{equation}

\begin{figure}
		\includegraphics[trim=410 0 410 0,clip,scale=0.2]{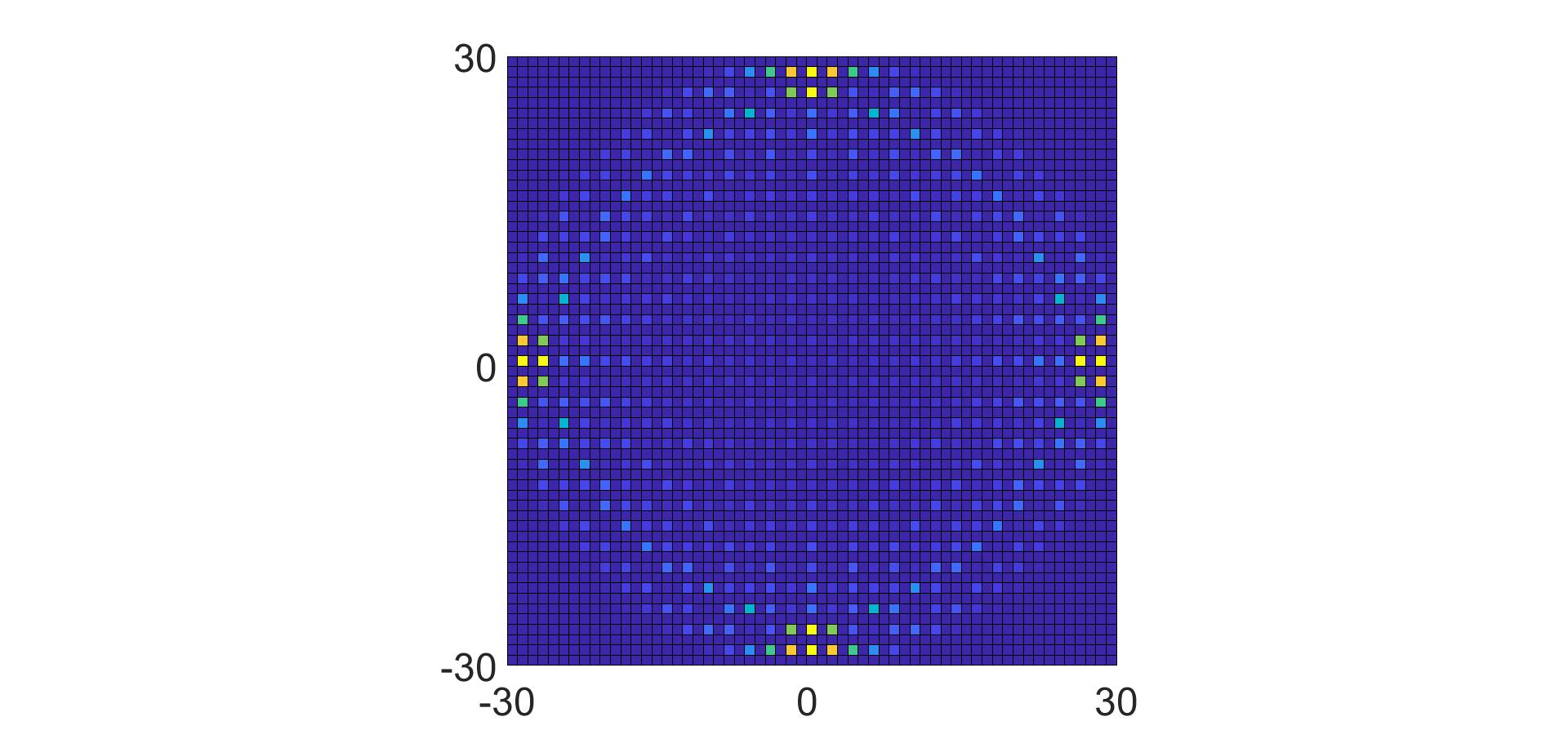}
		\caption{\label{fig:Grover}  Probability distribution of a Grover walk after $29$ steps with initial state of Eq.(\ref{eq:super_standmax}).}
	\end{figure}

	The probability of finding the walker on node $\left(x,y\right)$ at time $t$ is 
	\begin{equation}
	P\paren{x,y,t}= \sum_{c_x, c_y=0}^{1} \Mod{\psi_{c_xc_y}\paren{x,y,t}}^{2}.
	\end{equation}
	Fig.~\ref{fig:Grover} depicts the probability distribution at $t=29$ steps using the initial condition
	\begin{eqnarray}
	\label{eq:super_standmax}
	\ket{\psi\left(0\right)}=\frac{1}{2}\left(\ket{\nearrow} -\left|\searrow\right\rangle -\ket{\nwarrow} +\ket{\swarrow} \right)
	\otimes& \ket{00},
	\end{eqnarray}	
	which generates the largest standard deviation for the Grover coin~\cite{SPB15}.

	\section{Interacting quantum walk with HPP collision rules}\label{sec:model}

	There are many remarkable similarities between the classical HPP model and quantum walks on two-dimensional lattices. These similarities allow us to integrate these models in order to describe a version of  multiparticle quantum walks on the two-dimensional lattice. In this work we focus on the two-particle case.
	
	The Hilbert space of a two-particle quantum walk is
	\begin{equation}
	\label{eq:Hilbert_two}
	\mathcal{H}_{1}\otimes\mathcal{H}_{2},
	\end{equation}
	where $\mathcal{H}_1$ and $\mathcal{H}_2$ are the Hilbert spaces of the first and second walker, respectively, and each Hilbert space is a copy of the space in Eq.(\ref{eq:Hilbert_coined}). A generic state of the quantum walk at time $t$ is 
	\begin{equation}
	\label{eq:two_walkers}
	\left|\psi\left(t\right)\right\rangle =\sum_{c_{1},c_{2}}\sum_{l_{1},l_{2}}\psi_{c_{1}c_{2}}\left(l_{1},l_{2},t\right)\ket{c_{1},l_{1}} \otimes\ket{c_{2},l_{2}},
	\end{equation}
	where $\psi_{c_{1}c_{2}}\left(l_{1},l_{2},t\right)$ are the amplitudes at time $t$, $l_1 = (x_1,y_1)$ is the position of the first walker, and $c_1 = (c_{x_1},c_{y_1})$ is its coin state. Analogously, $l_2 = (x_2,y_2)$ is the position of the second walker and $c_2 = (c_{x_2},c_{y_2})$ is its coin state.
	In the non-interacting case, the evolution operator is $U_{1}\otimes U_{2}$, where  $U_1$ and $U_2$ are the evolution operators of the first and second walker, respectively, and each evolution operator is given by~(\ref{eq:coin_evolution}). 
	
	The evolution operator in the interacting case is defined as
	\begin{multline*}
	U_{\text{HPP}}\ket{c_{1},l_{1}} \ket{c_{2},l_{2}} = \nonumber\\
	\begin{cases}
	U_{\text{int}}\ket{c_{1},l_{1}} \ket{c_{2},l_{2}} & \text{if \ensuremath{l_1=l_2},}\\
	U_{1} \ket{c_{1},l_{1}} U_{2}\ket{c_{2},l_{2}}& \text{otherwise.}
	\end{cases}
	\end{multline*}
	When acting on basis states for which $(x_1,y_1)= (x_2,y_2)$, $U_{\text{HPP}}$ reduces to $U_{\text{int}}$ and, for the other cases, $U_{\text{HPP}}$ reduces to the non-interacting evolution operator $U_{1}\otimes U_{2}$. $U_{\text{int}}$ is defined as
	\begin{equation}
	\label{eq:v_int}
	U_{\text{int}}=\left(S_{1}\otimes S_{2}\right)\mathcal{C},
	\end{equation}
	where $S_{1}$ and $S_{2}$ are the shift operators of each walker and $\mathcal{C}$ is the collision operator, which acts on the total Hilbert space (\ref{eq:Hilbert_two}). The action of $\mathcal{C}$ on basis states such that $(x_1,y_1)=(x_2,y_2)$ is
	\begin{eqnarray}
	\label{eq:collision_pres}
	\mathcal{C}\ket{c_{x}c_{y}}\ket{xy} \otimes\ket{c'_{x}c'_{y}}\ket{xy} =\hspace{3cm}\\
	\begin{cases}
	\ket{\bar{c'}_{y}c'_{x}}\ket{xy} \otimes \ket{\bar{c}_{y}c_{x}}\ket{xy} & \text{if \ensuremath{(c'_{x}, c'_{y})=(\bar{c}_{x},\bar{c_{y}})}},\\
	\ket{c_{x}c_{y}}\ket{xy} \otimes\ket{c'_{x}c'_{y}}\ket{xy}  & \text{otherwise},
	\end{cases}\nonumber
	\end{eqnarray}
	where the position of both walkers is node $(x,y)$ and $\bar{c}\equiv c +1 \mod 2$. Note that the collision operator acts only when the position of the walkers coincides and its action does not change the position of the walkers. The condition $\ensuremath{(c'_{x}, c'_{y})=(\bar{c}_{x},\bar{c_{y}})}$ in~(\ref{eq:collision_pres}) implements the rule $\mbox{}_{\bullet}^{\circ}\mbox{}_{\circ}^{\bullet} \,\longleftrightarrow\, \mbox{}_{\circ}^{\bullet}\mbox{}_{\bullet}^{\circ}$ of the collision operator of the classical HPP model. 
	
	For head-on collisions, the direction of each walker rotates by $\pm 90^\circ$. For instance, if the coin state of the first walker is $\ket{00}$ before the collision, then the coin state after the collision is $\ket{01}$. Note that the notion of head-on collision in the quantum case is different from the classical notion because in the quantum case the state of the positions of the walkers is spread out. When we mention a head-on collision at a site, we have to consider that there is an amplitude associated with this collision.

	We have described the dynamics for the two-walker case and it is clear that the extension for three or more walkers is straightforward. However, the computational resources required to simulate the dynamics for the multiwalker case is huge because the number of dimensions of the Hilbert space is $4^{n}N^{2n}$ for a finite lattice with $N^2$ nodes and $n$ walkers.

	Most papers addressing interacting quantum walks employs the phase interaction~\cite{RLW15,SXBC15,WL16}, which is different from the one based on the HPP model. The evolution operator based on the phase interaction is defined as
	\begin{multline*} 
	U_{\text{phase}}\ket{c_{1},l_{1}} \ket{c_{2},l_{2}} = \nonumber\\
	\begin{cases}
	-U_{1} \ket{c_{1},l_{1}} U_{2}\ket{c_{2},l_{2}} & \text{if \ensuremath{l_1=l_2},}\\
	\,\,\,\,\,\, U_{1} \ket{c_{1},l_{1}} U_{2}\ket{c_{2},l_{2}}& \text{otherwise.}
	\end{cases}
	\end{multline*}
	This evolution operator changes the phase of the quantum state when the walkers are in the same position. This kind of dynamics was analyzed in the one-dimensional case in many papers such as~\cite{AAMSWW12,BW11,MolecularQW}.
	
	Despite the differences, both models obey symmetries~\cite{AAMSWW12} that can be used to simplify the simulations and calculations, e.g. the parity of each entry of the relative coordinate $(x_1,y_1)-(x_2,y_2)$, where $(x_1,y_1)$ and $(x_2,y_2)$ are the positions of the first and second walker, which follows from the fact that the shift operator changes $x$ and $y$ by $\pm 1$ in every step.


	\section{Numerical Results}\label{sec:numerics}

	To understand the dynamics, we focus on three quantities: 1)~The probability distribution of the position of the first walker, 2)~the standard deviation of the position of the first walker as a function of the number of steps, and 3)~the entanglement between the walkers as a function of the number of steps. We consider four localized initial states, three of them are separable and one of them has maximal entanglement between the coin systems. To perform the numerical calculations, we change the order of the subspaces because this simplifies the description of the initial states. From now on, we assume that the Hilbert space is spanned by the set of basis states $\ket{c_{x_1}c_{x_1}}\ket{c_{x_2}c_{x_2}}\ket{x_1y_1}\ket{x_2y_2}$, where  $x_1,y_1,x_2,y_2\in\mathbb{Z}$, $\ket{c_{x_1}c_{x_1}}$ and $\ket{c_{x_1}c_{x_1}}$ are the states of the coin of the first and second walker, respectively, and the definition of $U_{\text{HPP}}$ has been changed accordingly. We have selected the following initial states:
	\begin{align}
	\ket{\text{Sep}1} &=\frac{1}{2}\paren{\ket{\nearrow} -\ket{\searrow} -\ket{\nwarrow} +\ket{\swarrow}}\otimes\nonumber\\
	&\frac{1}{2}\paren{\ket{\nearrow} -\ket{\searrow} -\ket{\nwarrow} +\ket{\swarrow}} \ket{00}\ket{00} ,\label{eq:sep1}
	\end{align}
	\begin{align}
	\ket{\text{Sep}2} &=\frac{1}{2}\paren{\ket{\nearrow} -\ket{\searrow} -\ket{\nwarrow} +\ket{\swarrow}}\otimes\nonumber\\ 
	\frac{1}{2}&\paren{\ket{\nearrow} -\ket{\searrow} -\ket{\nwarrow} +\ket{\swarrow}}\ket{-1,-1}\ket{11} ,\label{eq:sep2}
	\end{align}
	\begin{align}
	\ket{\text{Grov}} &=\frac{1}{2}\paren{\ket{\nearrow} -\ket{\searrow} -\ket{\nwarrow} +\ket{\swarrow}}\otimes\nonumber\\
	\frac{1}{2}&\paren{\ket{\nearrow} -\ket{\searrow} -\ket{\nwarrow} +\ket{\swarrow}}\ket{00}\ket{-1,-1},\label{eq:Grover}
	\end{align}
	\begin{align}
	\ket{\text{Ent}} =\frac{1}{2} \big( & \ket{\nearrow}\ket{\nwarrow} + i\,\ket{\searrow} \ket{\swarrow} -\ket{\nwarrow} \ket{\nearrow} +\nonumber\\
	& i\,\ket{\swarrow} \ket{\searrow} \big)\ket{00}\ket{00}\label{eq:ent3}.
	\end{align}
	These initial states represent a broad spectrum of possibilities, which include (1)~entangled coins $(\ket{\text{Ent}})$ and non-entangled coins $(\ket{\text{Sep}1}, \ket{\text{Sep}2}, \ket{\text{Grov}})$, and (2)~same initial position $(\ket{\text{Sep}1}, \ket{\text{Ent}})$ and different initial positions $(\ket{\text{Sep}2}, \ket{\text{Grov}})$. The entanglement in the coin state of $\ket{\text{Ent}}$ is based on generalized Bell states~\cite{SL09}. On the other hand, the initial state $\ket{\text{Grov}}$ has non-entangled coins with walkers at neighboring nodes. This state represents non-interacting walkers each one with an initial state of the form of Eq.(\ref{eq:super_standmax}), which yields the maximum spreading rate for the mean square displacement for a single walker.

	
	\subsection{Probability distribution}
\begin{figure}
		\includegraphics[trim=460 0 360 0,clip,scale=0.14]{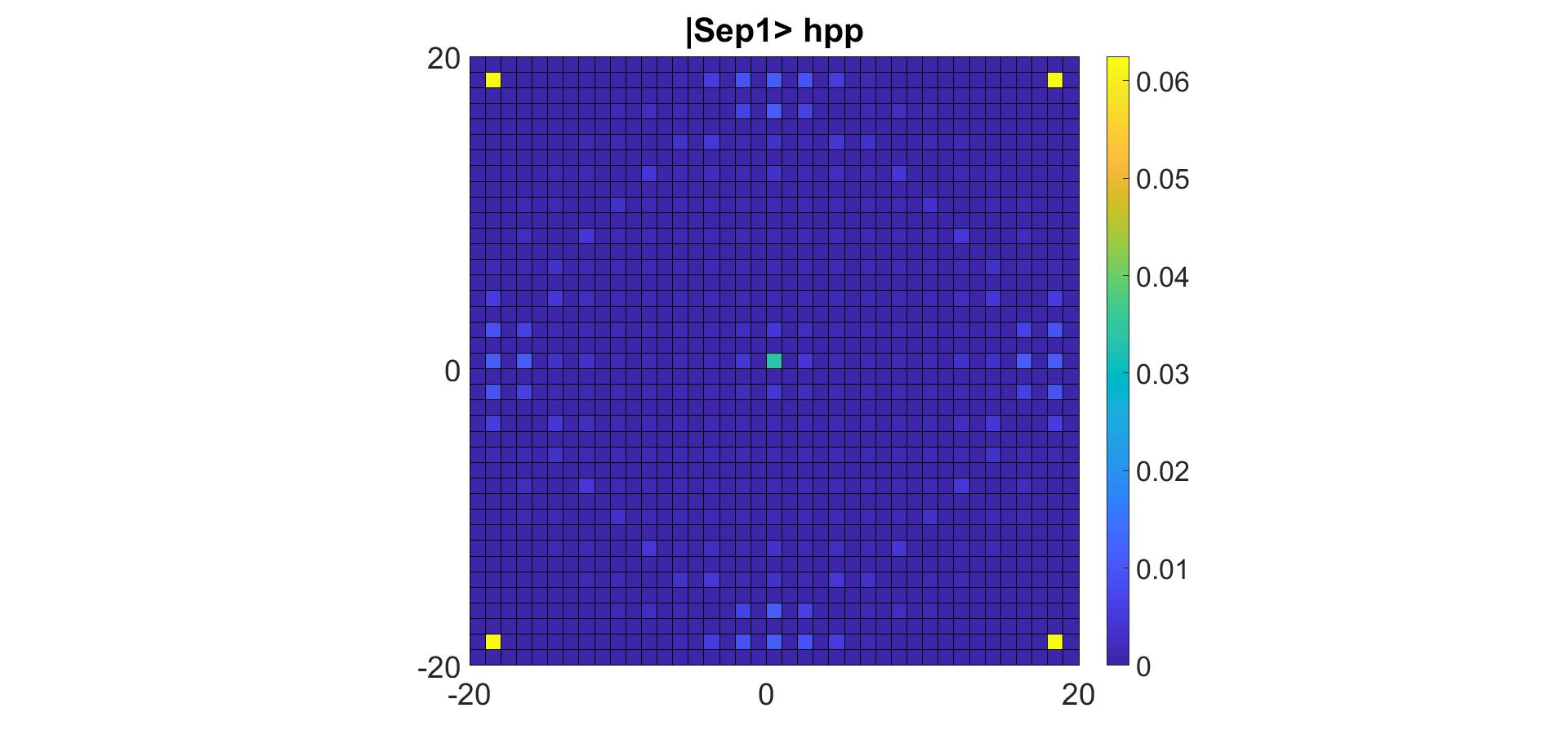}
		\includegraphics[trim=460 0 360 0,clip,scale=0.14]{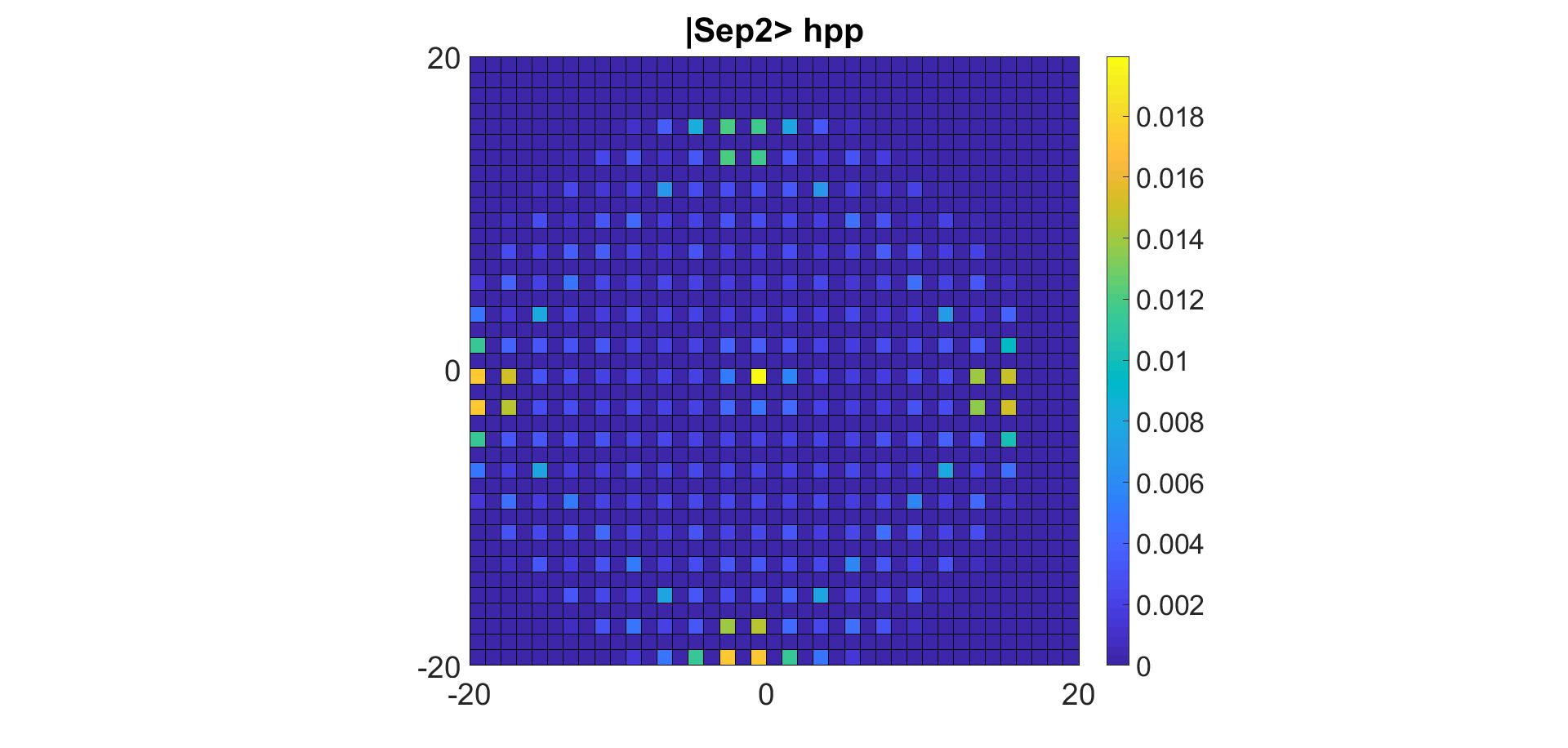}\\
		\includegraphics[trim=460 0 360 0,clip,scale=0.14]{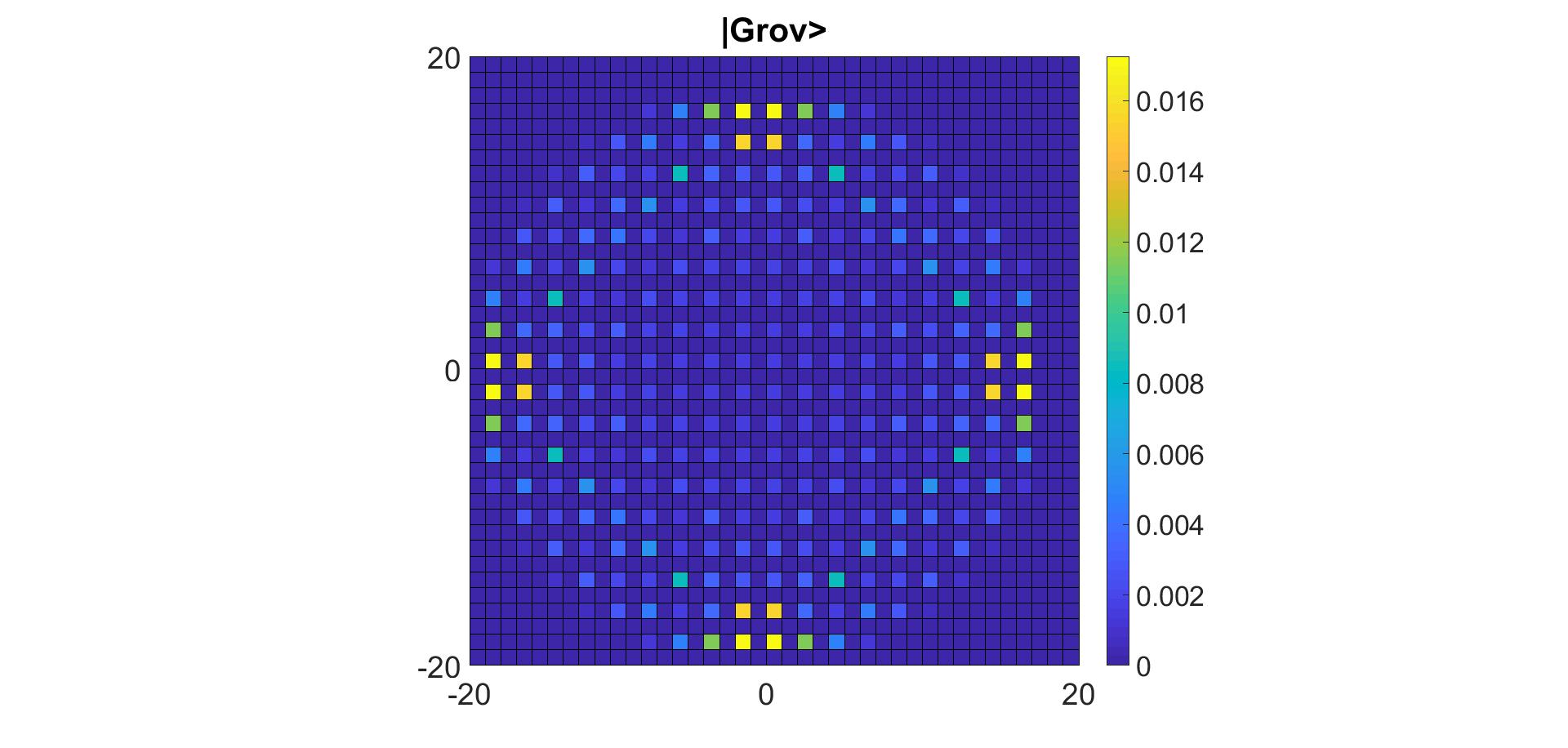}
		\includegraphics[trim=460 0 360 0,clip,scale=0.14]{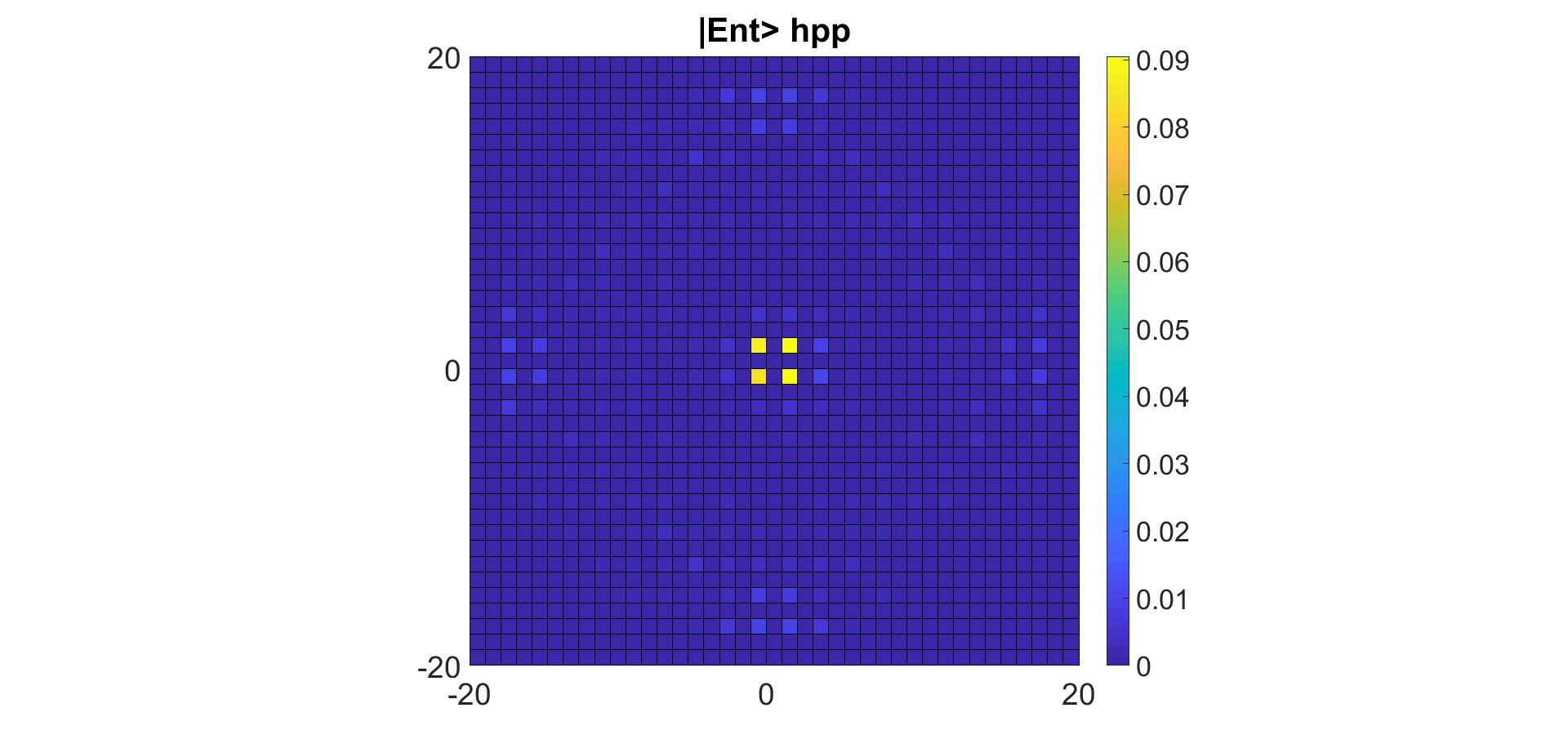}
		\caption{
			{\label{fig:allICs}  Probability distribution of the position of the first walker after $18$ steps with the \textit{HPP interaction} (the initial state is described in each panel).}
		}
	\end{figure}

	The probability distribution $P(x_1,y_1,t)$ associated with a measurement of the position of the first walker after $t$ steps is 
	\begin{eqnarray}\label{eq:P1stwalker}
	P\left(x_{1},y_{1},t\right)=
	\sum_{c_{x_{1}}c_{y_{1}}}\sum_{c_{x_{2}} c_{y_{2}}}\sum_{x_{2}\,y_{2}}\hspace{2cm}\nonumber\\
	\Big|\bracket{c_{x_{1}}c_{y_{1}}c_{x_{2}}c_{y_{2}}x_{1}y_{1}x_{2}y_{2}}{\psi\paren{t}} \Big|^2,
	\end{eqnarray}
	where  $\ket{\psi\paren{t}}=\left(U_{\text{HPP}}\right)^{t}\ket{\psi\paren{0}}$. Fig.~\ref{fig:allICs} depicts $P(x_1,y_1,t)$ when $t=18$ steps for each initial condition.

	For the initial condition $\ket{\text{Sep1}}$ (\ref{eq:sep1}), the probability distribution of the first walker after 18 steps is depicted in the first panel of Fig.~\ref{fig:allICs}. Initially, the position of both walkers is the origin and the coin state is separable (the initial state of each walker is given by~(\ref{eq:super_standmax}), which produces the largest spread in the one-walker case).  When we expand the tensor product of the coin states of the initial state, we obtain three sets of terms: 1)~four terms that have the same direction, such as $\ket{\nearrow}\ket{\nearrow}$, 2)~eight terms that have orthogonal directions, such as  $\ket{\nearrow} \ket{\searrow}$, and 3)~four terms with opposite directions (head-on collisions), such as $\ket{\nearrow} \ket{\swarrow}$. Since the walkers are initially in the same position, the action of the evolution operator is equal to the action of 
	$U_{\text{int}}=\left(S_{1}\otimes S_{2}\right){\mathcal{C}}$. 
	The action of $U_{\text{int}}$ on states of the first set, that is, on states such as $\ket{\nearrow}\ket{\nearrow}$, simply shifts both walkers in the same direction and the coin state of each walker remains the same. This explains the four yellow dots at the corners of the first panel of Fig.~\ref{fig:allICs}, which is a ballistic behavior with constant amplitude representing 1/4 of the total probability. The action of $U_{\text{int}}$ on states of the second set also acts trivially in their coin state then each walker keeps moving on the same direction. In the last set, we are in the case of head-on collision, thus   $U_{\text{int}}$ changes their directions. In both cases, the second and third sets, in the next step the action of the evolution operator produces a spreading via the coin operator, which is confirmed by the faded blued dots at the four cardinal direction (north, east, south, and west). Note, however, that the probability at the origin is relatively high suggesting the presence of localization.

	For the initial condition $\ket{\text{Sep2}}$ (\ref{eq:sep2}), the probability distribution of the first walker after 18 steps is depicted in the second panel of Fig.~\ref{fig:allICs}. Initially, the positions of the walkers are different (the distance is even) and the coin state is separable (equal to the coin state of $\ket{\text{Sep1}}$). When we expand the tensor product of the coin states of the initial state, we obtain the same sets of terms described before. Since the walkers are in different positions (no head-on collision), the action of the evolution operator spreads the positions of the walkers following the quantum-walk dynamics. In the next step there is the presence of head-on collisions. For instance, the action of the evolution operator on the term $\ket{\nearrow} \ket{\swarrow}\ket{-1,-1}\ket{11}$ produces a nonzero amplitude for the term $\ket{\nearrow} \ket{\swarrow}\ket{00}\ket{00}$, which represents a head-on collision. In the next steps, there are the superpositions of more and more head-on collisions, and the probability distribution depicted in the second panel of Fig.~\ref{fig:allICs} shows the effect of the interaction on the first walker. Note that the probability at the origin is high compared to the one at the origin of third panel, suggesting the presence of localization produced by the interaction.

	For the initial condition $\ket{\text{Grov}}$ (\ref{eq:Grover}), the probability distribution of the first walker after 18 steps is depicted in the third panel of Fig.~\ref{fig:allICs}. Initially, the positions of the walkers are different (the distance is odd) and the coin state is separable (equal to the coin state of $\ket{\text{Sep1}}$). Since the initial distance between the walkers is odd, there is no head-on collision at any time step. This means that the dynamics of the walkers is decoupled at any time step, and the probability distribution depicted in the third panel of Fig.~\ref{fig:allICs} coincides with the probability distribution of the Grover walk. We can turn off the interaction by choosing an appropriate initial condition.
	

	For the initial condition $\ket{\text{Ent}}$ (\ref{eq:ent3}), the probability distribution of the first walker after 18 steps is depicted in the fourth panel of Fig.~\ref{fig:allICs}. Initially, the position of both walkers is the origin and the coin states of the walkers are entangled. Later, part of the wave function spreads as we can see from the faded blue dots at the cardinal directions while another part remains localized near the origin as we can see from the yellow dots at the nearest nodes to the origin. The wave function oscillates between the origin and the nearest nodes.
	\begin{figure}
		\includegraphics[trim=460 0 360 0,clip,scale=0.14]{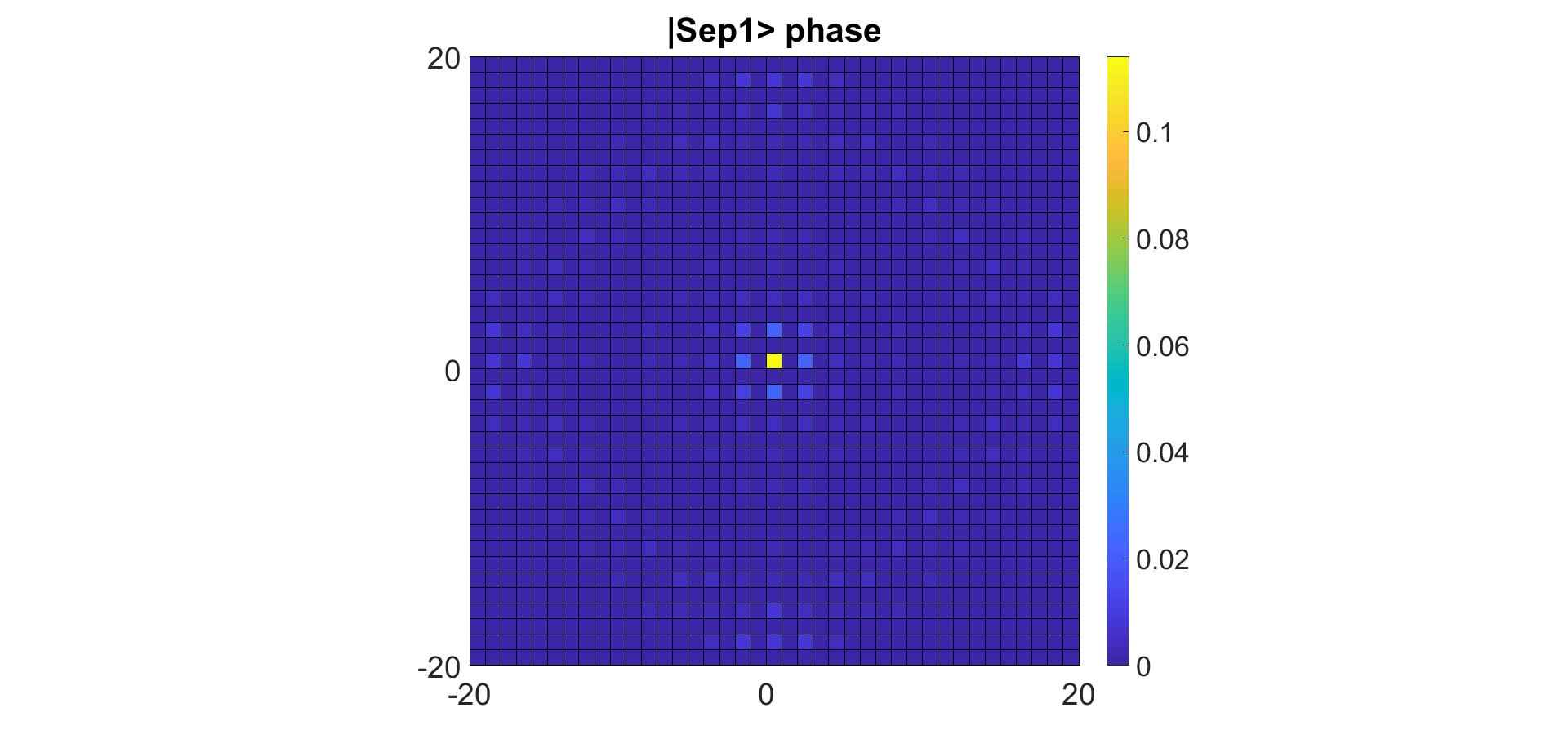}
		\includegraphics[trim=460 0 360 0,clip,scale=0.14]{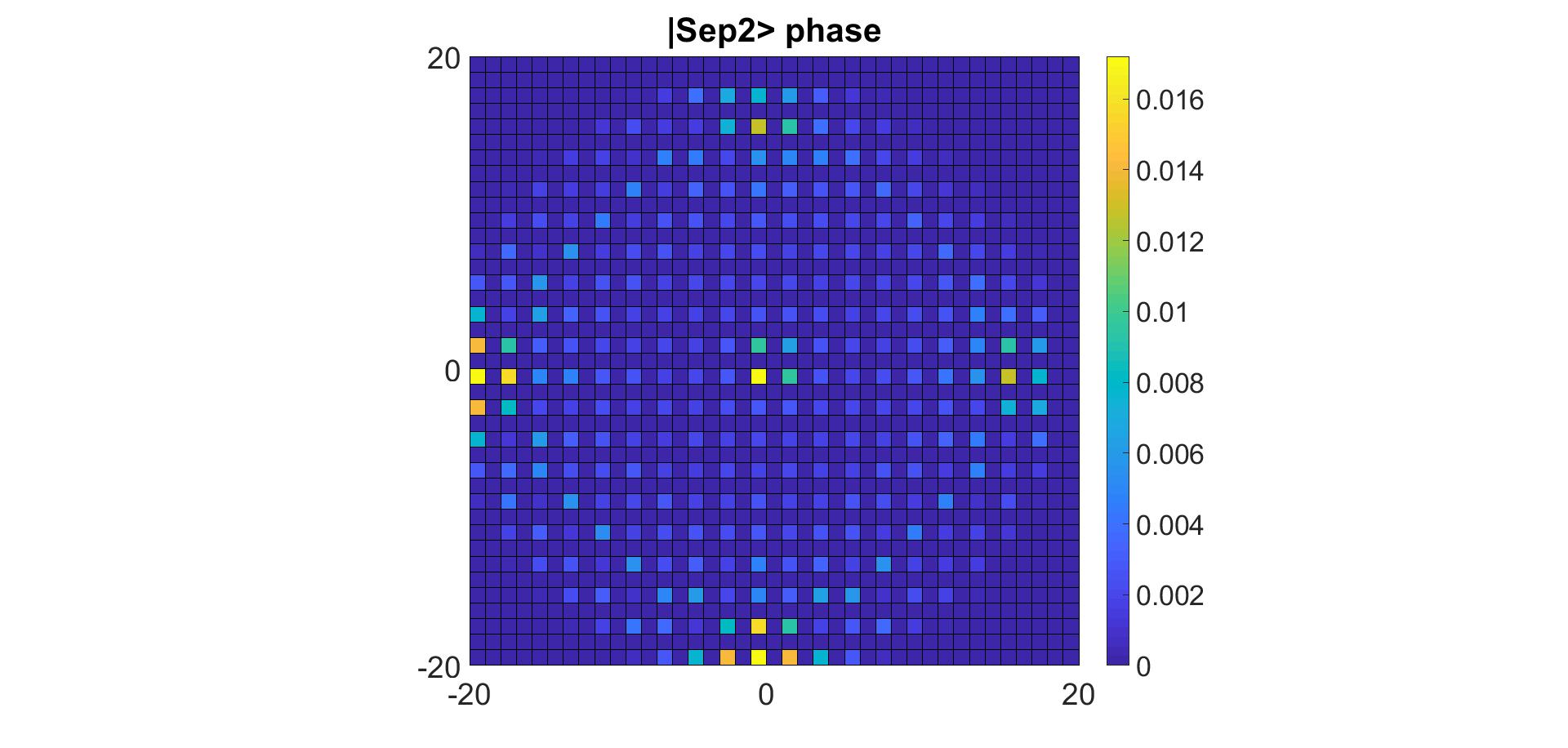}\\
		\includegraphics[trim=460 0 360 0,clip,scale=0.14]{Grov}
		\includegraphics[trim=460 0 360 0,clip,scale=0.14]{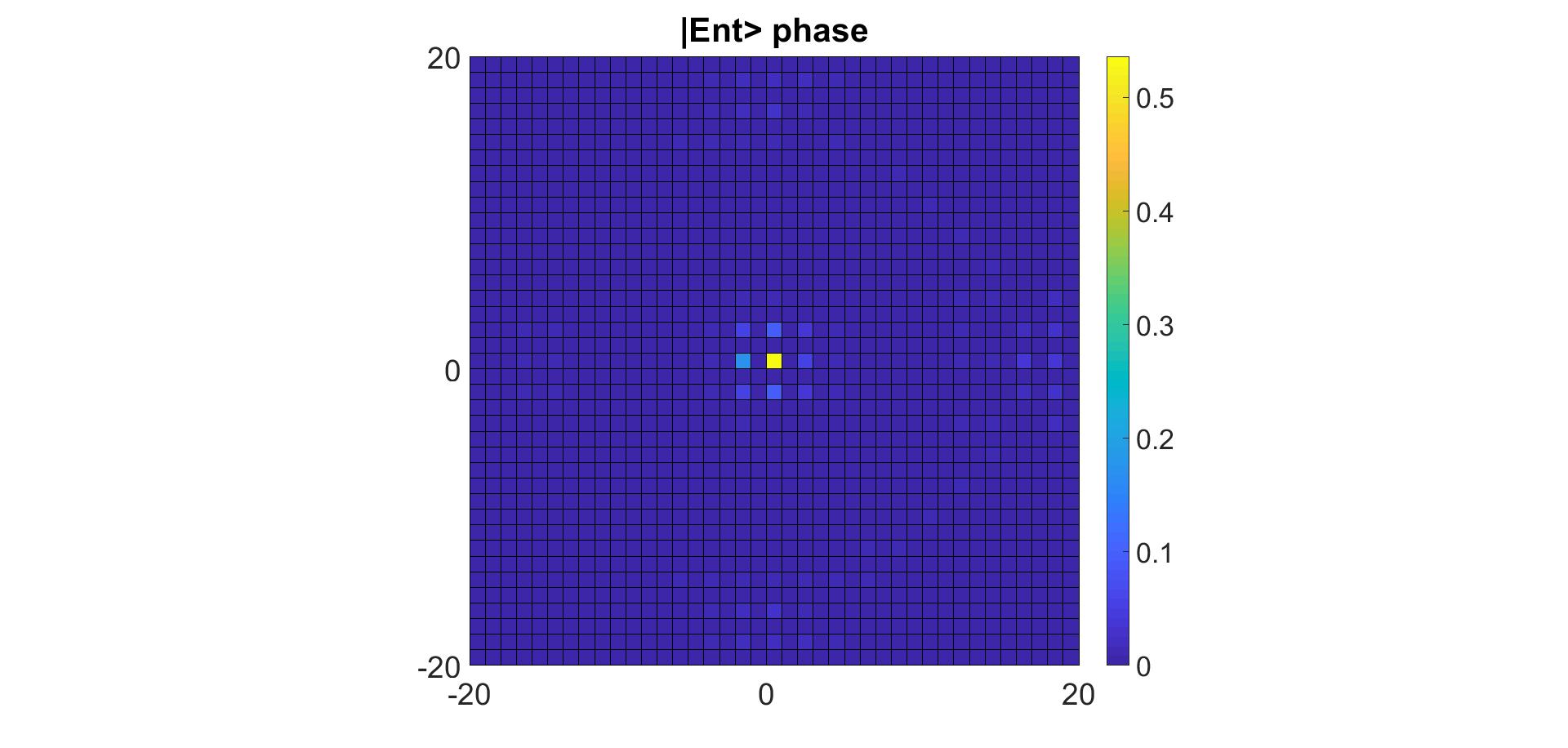}
		\caption{
			{\label{fig:allICphase}   Probability distribution of the position of the first walker after $18$ steps with the \textit{phase interaction} (the initial state is described in each panel).}
		}
	\end{figure}

	Fig.~\ref{fig:allICphase} depicts $P(x_1,y_1,t)$ when $t=18$ steps for the same initial conditions analyzed in Fig.~\ref{fig:allICs} using the evolution operator $U_\text{phase}$ (instead of $U_\text{HPP}$).
	The probability distributions produced by the phase interaction are similar to the probability distributions produced by the HPP interaction. For the initial condition $\ket{\text{Grov}}$, the results (third panels) are exactly the same because the interaction is turned off. For the other initial conditions, we note that the probability distribution spreads a bit more in the HPP-interaction case and the localization at the origin is a bit larger for the phase-interaction case. In the next sections, we obtain quantitative results that help to explain this behavior.
	
	\subsection{Standard deviation}
	
\begin{figure}
		\includegraphics[trim=58 0 50 0,clip,scale=0.28]{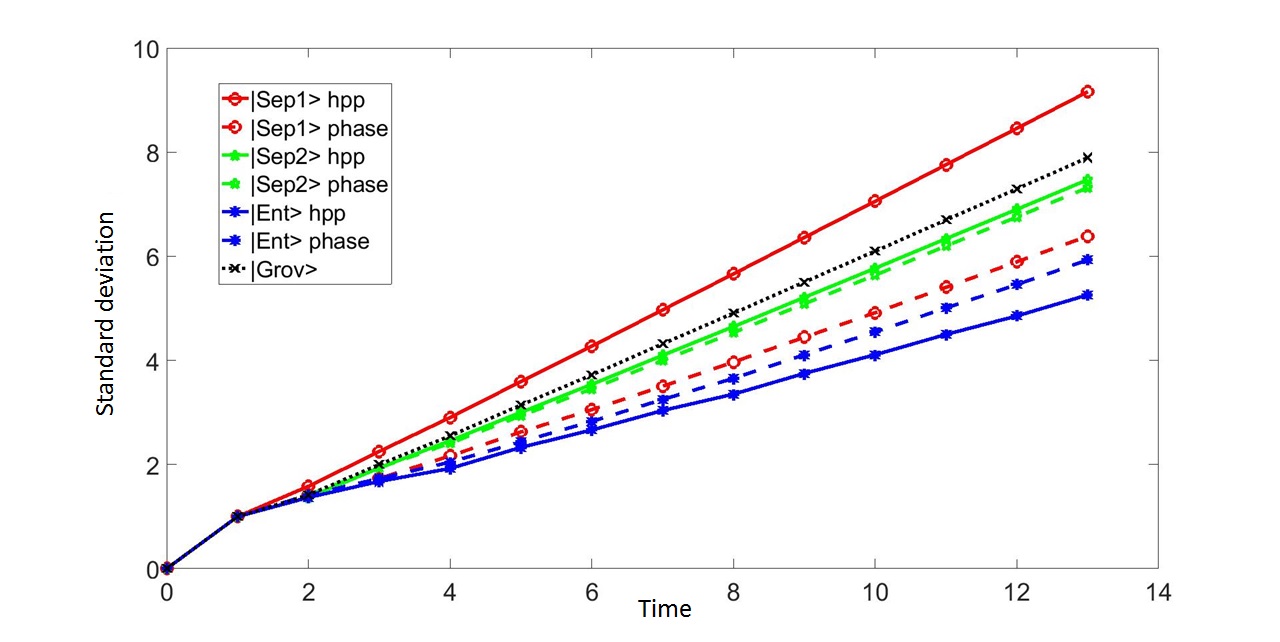}
		\caption{\label{fig:stand} Standard deviation of the position of the first walker as a function of the number of steps for the initial conditions described at the upper-left corner. Continuous lines refer to $U_\text{HPP}$ and dashed lines to $U_\text{phase}$.}
	\end{figure}

	The standard deviation of the position of the first walker is
	\begin{equation*}
	\sigma(t)=\sqrt{\sum_{x_1y_1}P(x_1,y_1,t)\,\big((x_1-\bar{x}_1)^{2}+(y_1-\bar{y}_1)^{2}\big)},
	\end{equation*} 
	where $P(x_1,y_1,t)$ is given by~(\ref{eq:P1stwalker}) and
	\begin{eqnarray*}
		\bar{x}_1&=&\sum_{x_1y_1}x_1P(x_1,y_1,t),
		\\\bar{y}_1&=&\sum_{x_1y_1}y_1P(x_1,y_1,t).
	\end{eqnarray*}  
	Fig.~\ref{fig:stand} depicts $\sigma(t)$ as a function of the number of steps for the initial conditions (\ref{eq:sep1})-(\ref{eq:ent3}). Continuous lines are used for the evolution driven by $U_\text{HPP}$ and dashed lines for the evolution driven by $U_\text{phase}$. The figure shows that $\sigma(t)=\alpha\,t$ for sufficiently large $t$, where the slope $\alpha$ depends on the initial state and $0<\alpha<1$.
	
	The largest slope is attained with the initial condition $\ket{\text{Sep1}}$ and the HPP interaction. As we can see in the first panel of Fig.~\ref{fig:allICs}, there are four yellow dots at the corners of the figure, which give a large contribution to the standard deviation. The slope of $\sigma(t)$ as a function of $t$ can be further increased (compared to the slope produced when the initial state is $\ket{\text{Sep1}}$) by taking a new initial state such that the amplitudes of the terms $\ket{\nearrow}\ket{\nearrow}$, $\ket{\searrow}\ket{\searrow}$, $\ket{\nwarrow}\ket{\nwarrow}$, $\ket{\swarrow}\ket{\swarrow}$ are larger. The slope of $\sigma(t)$ when the evolution is driven by $U_\text{phase}$ is considerably smaller for the initial condition $\ket{\text{Sep1}}$. We explain this fact by pinpointing that the coin operator spreads the coin state in the phase-interaction case even when the walkers are in the same position, that is, the amplitudes of the terms $\ket{\nearrow}\ket{\nearrow}$, $\ket{\searrow}\ket{\searrow}$, $\ket{\nwarrow}\ket{\nwarrow}$, $\ket{\swarrow}\ket{\swarrow}$ decrease preventing the appearance of the four yellow dots at the corners of probability distribution as the ones we see the first panel of Fig.~\ref{fig:allICs}.

	The smallest slope is attained with the initial condition $\ket{\text{Ent}}$ and the HPP interaction. As we can see in the fourth panel of Fig.~\ref{fig:allICs}, the probability distribution is highly concentrated at the origin and the four yellow dots at the corners are absent. Localization means that part of the wave function does not spread, decreasing the standard deviation. This analysis highlights the qualitative features that play the key roles in quantifying the standard deviation and it can be extended to other initial conditions and to the phase interaction case in order to match the slopes of $\sigma(t)$ with the corresponding probability distribution.

	
	\subsection{Entanglement between the walkers}
	\label{sec:Ent}

	In this subsection, we analyze the entanglement between the walkers. As entanglement measure, we use the von Neumann entropy 
	\begin{equation}
	\label{eq:von_pho1}
	S(\rho_{1}(t))=-\text{Tr}\big(\rho_{1}(t)\log_2\rho_{1}(t)\big),
	\end{equation}
	where, for $\ket{\psi(t)}$ given by (\ref{eq:two_walkers}),
	\[
	\rho_{1}(t)=\text{Tr}_{2}\big(\ket{\psi(t)}\bra{ \psi(t)}\big).
	\]
	$\text{Tr}_2$ is the partial trace over the system of the second walker. The reduced density matrix of the first walker $\rho_1(t)$ acts on $\mathcal{H}_{1}=\mathcal{H}_{C}\otimes\mathcal{H}_{P}$.  The entropy of the second walker is equal to the entropy of the first walk, that is, $S\left(\rho_{2}(t)\right)=S\left(\rho_{1}(t)\right)$ because the total system is in a pure bipartite state. 
	
	\begin{figure}
		\includegraphics[trim=60 0 30 0,clip,scale=0.28]{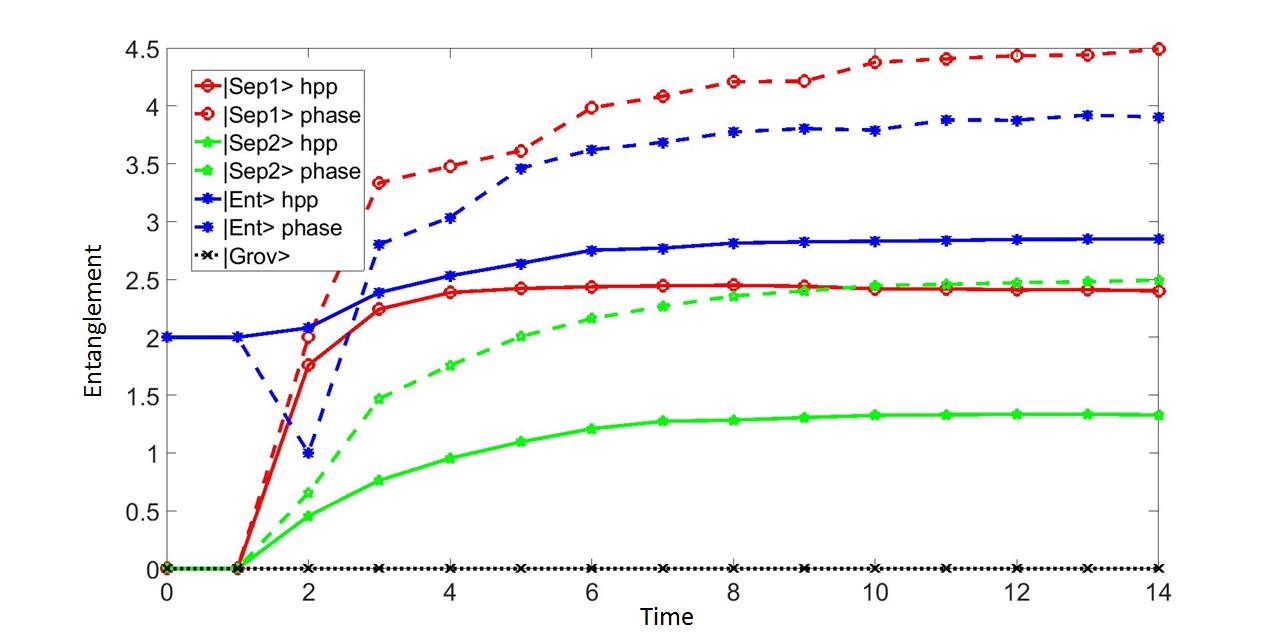}
		\caption{\label{fig:ent} The von Neumann entropy of the reduced density matrix of the first walker as a funtion of the number of steps for the initial conditions described at the lower-right corner. Continuous lines refer to $U_\text{HPP}$ and dashed lines to $U_\text{phase}$.}
	\end{figure}
	
	Fig.~\ref{fig:ent} depicts $S(\rho_{1}(t))$ as a function of the number of steps for the initial conditions (\ref{eq:sep1})-(\ref{eq:ent3}). We use again the convention that continuous lines describe the entanglement when the evolution is driven by $U_\text{HPP}$ and dashed lines describe the entanglement when the evolution is driven by $U_\text{phase}$. Entanglement is initially zero for the initial conditions $\ket{\text{Sep1}}$, $\ket{\text{Sep2}}$, and $\ket{\text{Grov}}$, which are separable states, and is nonzero for the initial condition $\ket{\text{Ent}}$, which is an entangled state. When the number of steps increases, the entanglement for all initial states quickly increase and tend to a constant value, as we can see from the plateaus of Fig.~\ref{fig:ent}.
	
	Let us explain why $S(\rho_{1}(t))$ tends to a constant value for large $t$ for the HPP interaction. Entanglement is produced only by the interaction between the walkers. The interaction occurs only if the decomposition of the state vector in terms of the computational basis has terms $\ket{c_{x_1}c_{x_1}}\ket{c_{x_2}c_{x_2}}\ket{x_1y_1}\ket{x_2y_2}$ such that the position of the walkers coincide, that is, $(x_1,y_1)= (x_2,y_2)$, and if their coins states are the ones that yield the head-on collision. We have to assess the contribution of those terms relative to the terms such that the position of the walkers do not coincide. Consider a 4-dimensional space with coordinates $x_1$, $y_1$, $x_2$ and $y_2$. After $t$ steps, the number of points such that $(x_1,y_1)\neq (x_2,y_2)$ is $O(t^4)$ while the number of points such that $(x_1,y_1)= (x_2,y_2)$ is $O(t^2)$. When the initial conditions are localized and the wave function is spreading, 
	both kind of terms have nonzero amplitudes, but the ratio between the number of terms with nonzero amplitude such that the position of the walkers coincide and the number of terms with nonzero amplitude such that the position of the walkers do not coincide goes to zero when the number of steps increase. This means that the source of entropy dwindles and $S(\rho_{1}(t))$ tends to a constant value.

	Note that, when we fix one of the initial conditions, which are given by Eqs.~(\ref{eq:sep1}) to~(\ref{eq:ent3}), $S(\rho_{1}(t))$ tends to a larger constant for the phase interaction compared to the HPP interaction. This seems to be a general pattern for localized initial conditions and we explain this behavior in the following way. Since $\sigma(t)$ is proportional to $t$, the wave function spreads more and more when the number of time steps increases. We can ignore localization is this argumentation. Consider the decomposition of the state vector in terms of the computational basis. The contribution of the amplitude of terms $\ket{c_{x_1}c_{x_1}}\ket{c_{x_2}c_{x_2}}\ket{x_1y_1}\ket{x_2y_2}$ such that $\sqrt{|x_1|^2+|y_1|^2}\sim \sigma(t)$ and $\sqrt{|x_2|^2+|y_2|^2}\sim \sigma_2(t)$, where $\sigma_2(t)$ is the standard deviation of the second walker, increases relative to the remaining terms when $t$ increases. The terms  $\ket{c_{x_1}c_{x_1}}\ket{c_{x_2}c_{x_2}}\ket{x_1y_1}\ket{x_2y_2}$ such that the corresponding momenta are pointing to the same outward direction have the largest contribution because they coincide with the direction of the spreading. This argument shows that the HPP interaction tends to fade for large $t$. The phase interaction continues to be active for part of those terms (the ones that have the same position), generating more entropy relative to the HPP interaction. In the end, Fig.~\ref{fig:ent} shows that the entropy production of both interactions tend to zero for large $t$.

	\

	\section{Conclusions}
	\label{sec:conc}
	
	We have introduced a quantum walk model with interacting walkers using the HPP collision rules. The HPP interaction is interesting from the viewpoint of physics and in the context of quantum walks because it is local, conserve momentum and energy, and may be useful to simulate quantum gases and complex quantum systems.
	
	We have defined the dynamics of the model by using a time evolution operator that implements the HPP collision rules when the walkers are on the same node and the standard quantum-walk evolution operator when the walkers are on different nodes. Our results were obtained by simulating two walkers on a two-dimensional lattice. We have analyzed numerically three quantities: (1)~The probability distribution of the position of one walker, (2)~the standard deviation of the position, and (3)~the entanglement between the walkers as a function of time. 
	The results suggest that this model might be a good candidate to be employed in quantum-walk-based search algorithms on the lattice, since the walker can spread faster with this type of interaction than with the phase interaction. It also spreads faster than a single quantum walker.
	
	Multiparticle quantum walks have been successfully used to analyze graph isomorphism and to prove that quantum walks are universal for quantum computation. It has the potential to be applied in many computational problems, such as spatial search algorithms, and to simulate the behavior of complex quantum systems, such as quantum gases. It is known that the classical HPP model cannot simulate the motion of viscous fluids described by the Navier-Stokes equations. The quantum version may inherit this limitation. A possible escape route in this case is to employ the interactions described in the Frisch, Hasslacher, and Pomeau (FHP) model~\cite{FHP86}. We intend address this issue in future publications. We also intend to address the phase interaction of two quantum walkers on a two-dimensional lattice with the goal of obtaining molecular states, similar to the one described in~\cite{MolecularQW}, which addressed interacting quantum walks on the line. 

\bibliographystyle{unsrt}

\end{document}